# AN OPEN-CHANEL MICROFLUIDIC MEMBRANE DEVICE FOR *IN SITU* HYPERSPECTRAL MAPPING OF ENZYMATIC CELLULOSE HYDROLYSIS

H.Y.N. Holman[1*], W. Zhao[1], J.D. Nill[2], L. Chen[1], S.R. Narayansamy[1], and T. Jeoh[2]

[1]Molecular Biophysics and Integrated Bioimaging, Lawrence Berkeley National Laboratory, California,
[2]Biological and Agricultural Engineering, University of California, Davis, California

## ABSTRACT

Synchrotron infrared hyperspectral microscopy is a label-free and non-invasive technique well suited for imaging of chemical events *in situ*. It can track the spatial and temporal distributions of molecules of interests in a specimen in its native state by the molecule's characteristic vibrational modes. Despite tremendous progress made in recent years, IR hyperspectral imaging of chemical events in biomaterials in liquids remains challenging because of the demanding requirements on environmental control and strong infrared absorption of water[1-2]. Here we report a multi-phase capillary-driven membrane device for label-free and real-time investigation of enzymatic deconstruction of algal cellulose purified from *Cladophora aegagropila*.

## INTRODUCTION

Commercialization of advanced biofuels hinges on efficient hydrolysis to release glucose from the cellulose fraction of lignocellulosic plant biomass which is then fermented to bioethanol or other biofuels and bioproducts. Cellulose functions as a structural polysaccharide in plant cell walls and resists enzymatic hydrolysis by packing into highly ordered (crystalline) fibrils that limits enzyme access to the glycosidic bonds (C−O−C). A fundamental understanding of the reaction mechanisms by which cellulase hydrolyzes cellulose remain unsolved[3], due in part to limitations of imaging capabilities to capture at micro-to-nano-scale structural and chemical information in real-time and *in situ*. Herein, we develop a novel integrated open-channel platform for label-free tracking of the location and kinetics of cellulose hydrolysis. Two key features include (1) a multi-layer multi-phase open-channel membrane device that controls moisture, temperature, and enzyme-solution flow, (2) long-term continuous measurements of C-O stretches of the glucose residues and the C-O-C of the cellulose. As a proof-of-concept for profiling cellulose deconstruction at a molecular-structural level, we demonstrated the temporally- and spatially-resolved chemical modification of cellulose imparted by a cellobiohydrolase enzyme Cel7A.

## EXPERIMENTAL

The polydiemthylsilonxane (PDMS) device (**Figure 1**) composed of a gold-coated silicon nitride porous membrane between a feeding channel and a humidity controlled viewing chamber. It allows cellulose to be attached and maintained on the upper membrane surface in a micrometer-thick layer of enzyme solution while buffer media is replenished from the feeding microchannels below. The humidity sensor is placed next to an observation window to monitor humidity in the humidity control chamber that uses vapor-based capillary-driven microfluidic regulation. For enhanced signal-to-noise detection, we use an infrared or atomic force microscope to focus an infrared beam, resulting in studying hydrolysis with a micrometer-to-nanometer spatial resolution.

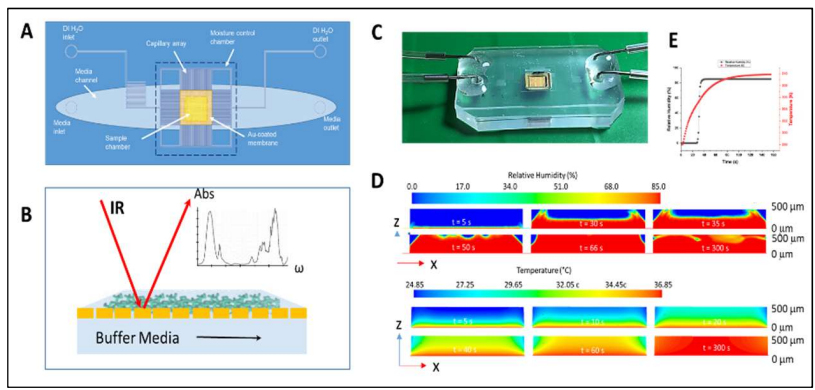

**Figure 1: A.** Microfluidic membrane device schematic. **B**. Cellulose fibrils are maintained in thin film of Cel7A-medium fluid that allows probing with IR light. Soluble enzyme is supplied from the media channel below the membrane. **C.** Device. **D, E**. Results of Computational Fluid Dynamic (CFD) simulation of the temperature and relative humidity fields for the device which are consistent with values from the temperature and humidity sensors.

## RESULTS AND DISCUSSION

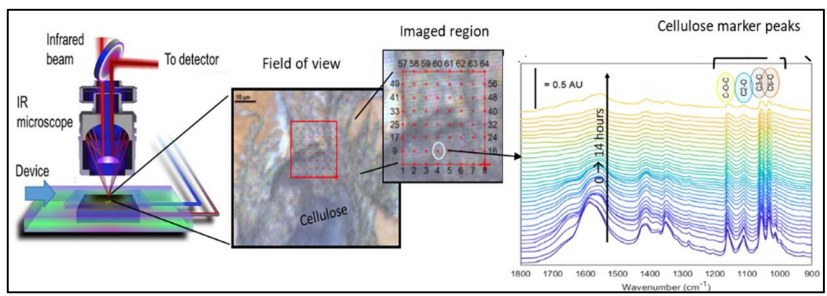

**Figure 2. Imaging of cellulose during enzymatic hydrolysis.** Continuous live measurements of hydrolysis were performed using synchrotron infrared (SIR)[2] illumination, using a 8x8 grid with 2.5 μm step size and 20 minute spacing for 14 hours. SIR spectra were obtained using transflection measurement mode between 4000 and 650 cm$^{-1}$ at 4 cm$^{-1}$ spectral resolution, and averaged 16 scans.

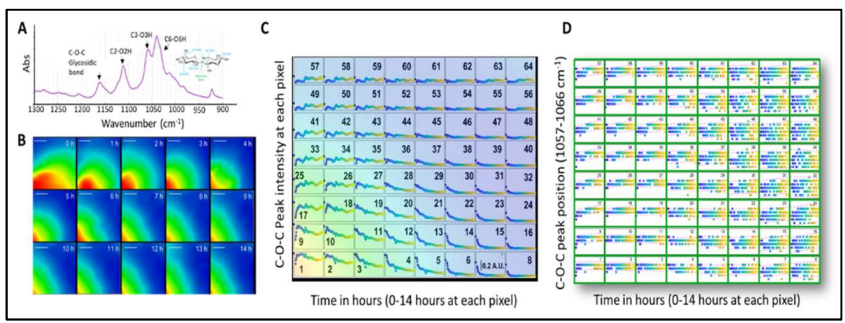

**Figure 3. Visualization of spatially resolved deconstruction of cellulose during hydrolysis. A.** IR spectrum of C. aegagropila cellulose and their diagnostic markers. **B.** Intensity heat map of glycosidic bond shows the cellulose "particle" shrinks and thins from outer to inner regions (imaged region in Figure 2). **C, D.** Tracking the location-specific kinetics of cellulose depletion and impact of hydrolysis on molecular ordering in cellulose.

Cellulose hydrolysis study was performed (**Figure 2**) and the optimum flow rate ratio for the two fluid inputs (**Figure 1**) was 1:2.5 (DI water in the upper channels: Cel7A-buffer medium in the lower channels) to maintain a micrometer thick liquid film and pH during live infrared imaging of hydrolysis. Algal cellulose fibrils purified from *C. aegagropila* were deposited onto the porous membrane and incubated overnight in the refrigerator for attachment. Prior to use, unbound cellulose was rinsed off the surface with enzyme-free buffer. At the start of the experiment, enzyme (*Trichoderma reesei* Cel7A (*Tr*Cel7A)) at a concentration of 3.5 μg/mL in sodium acetate pH 5 was injected into the bottom media channel at a flow rate of 0.5 μL/min. DI water was injected into the upper channel to feed the capillaries in the humidity control chamber at a flow rate of 0.2 μL/min to maintain a relative atmospheric humidity of ~85% at 37 °C. As a polymer of glucose, cellulose features C-O bonds at five of the six carbons of the glucose residues that contribute to absorption peaks (**Figure 3A**) prominently at 1034 cm$^{-1}$, 1060 cm$^{-1}$, and 1160 cm$^{-1}$, corresponding to C-O stretches at the sixth (C6) and third carbon (C3) of the glucose residues, and the glycosidic bond (C-O-C) of the polysaccharide. We documented spatially heterogeneous and temporal changes in the abundance (**Figure 3B, 3C**) and molecular ordering (**Figure 3D**) of cellulose during enzymatic hydrolysis over 12 hours. We observed a decline in absorbance intensity of the 1160 cm$^{-1}$ glycosidic bond peak, and wavenumber shifts in the prominent 1034 cm$^{-1}$, 1060 cm$^{-1}$, and 1160 cm$^{-1}$ peaks. In some regions of cellulose, 'blue shift' of these three peaks indicated that cellobiohydrolase action removed 'less ordered' cellulose, leaving intact the intra-and intermolecular hydrogen bonding by C3 and C6 in cellulose. In regions of thick cellulose, 'red shift' of these same peaks indicated disordering, likely due to active decrystallization of cellulose by enzyme complexation and hydrolysis.

## CONCLUSION

This work demonstrates for the first time the capability of *in situ* IR imaging for profiling the spatial and temporal changes in cellulose molecular ordering and deconstruction during enzymatic hydrolysis by coupling infrared microscopy with open-channel temperature and humidity controlled microfluidics membrane devices.

## ACKNOWLEDGEMENTS

This work was supported by the U.S. DOE, Office of Biological and Environmental Research (BER),

**CONTACT:** H.Y.N. Holman; phone +1-510-486-5943; hyholman@lbl.gov; Website: https://bsisb.lbl.gov/